\documentclass[aps,prd,showpacs, nofootinbib,11pt]{revtex4-1}
\usepackage{latexsym,bm,amsmath,amssymb,amsfonts}

\usepackage{graphicx,shortvrb}
\usepackage{epsfig}
\usepackage{newlfont}
\usepackage{hyperref}
\usepackage{xcolor}
\usepackage{cool}
\hypersetup{
    colorlinks,
    linkcolor={red!50!black},
    citecolor={blue!50!black},
    urlcolor={blue!80!black}
}

\begin{document}

\title{Kundt solutions of Minimal Massive 3D Gravity}

\author{Nihat Sadik Deger}
\email{sadik.deger@boun.edu.tr}
\affiliation{Department of Mathematics, 
Bo\u{g}azi\c{c}i University, 34342, Bebek, Istanbul, Turkey}
\author{{\"O}zg{\"u}r Sar{\i}o\u{g}lu}
\email{sarioglu@metu.edu.tr}
\affiliation{Department of Physics, Faculty of Arts and  Sciences,\\
              Middle East Technical University, 06800, Ankara, Turkey}

\date{\today}

\begin{abstract}
We construct Kundt solutions of Minimal Massive Gravity theory and show that, similar to 
Topologically Massive Gravity (TMG), most of them are constant scalar invariant (CSI) spacetimes,
that correspond to deformations of round and warped (A)dS. We also find an explicit non-CSI 
Kundt solution at the merger point. Finally, we give their algebraic classification with respect 
to the traceless Ricci tensor (Segre classification) and show that their Segre-types match with the 
types of their counterparts in TMG.
\end{abstract}


\maketitle

\section{Introduction}\label{intro}
In principle, the celebrated AdS/CFT duality can be used in giving a `quantum description' of
a gravity theory possessing an anti de-Sitter (AdS) vacuum, or at the very least, in ruling out
the existence of such a `consistent' quantum theory. Yet, the construction of a physically
viable theory that passes this powerful test has been a big challenge. Since two-dimensional
conformal field theories (CFTs) are better understood, three-dimensional gravitational models 
are in a better shape for the application of this duality. One such 3D model is the `good old'
cosmological ``Topologically Massive Gravity'' (TMG) \cite{Deser:1982vy, Deser:1981wh}, where 
the `opposite sign' Einstein-Hilbert action with a cosmological constant is modified by the addition 
of the parity-odd gravitational Chern-Simons term. TMG has a a single massive propagating 
graviton mode, and when the cosmological constant is negative it has an AdS vacuum. The 
renowned BTZ black hole \cite{Banados:1992wn} is also a solution to TMG, which makes 
the theory even more interesting. Unfortunately, the central charge of the dual CFT is negative 
when the bulk graviton has positive energy, and this destroys unitarity. This, so-called ``bulk vs
boundary clash,'' problem persists also in the parity-preserving ``New Massive Gravity'' (NMG) 
model \cite{Bergshoeff:2009aq, Bergshoeff:2009hq}, which has an action described by the 
addition of a particular combination of curvature-squared terms to the usual Einstein-Hilbert term.

Remarkably, a `deformation' of TMG theory, recently proposed in \cite{Bergshoeff:2014pca} and 
termed as ``Minimal Massive Gravity" (MMG), solves the `bulk vs boundary clash' problem for
a certain range of its parameters. This theory is `minimal' in the sense that it has only one
massive spin-2 mode in the bulk, i.e. the same minimal local structure as TMG. These features
considerably promote MMG and determining its exact solutions is quite an interesting problem
on its own. The exact solutions of MMG, that have been found so far, are as follows: Apart from
static black hole solutions, which obviously exist for specific values of parameters, and an
(A)dS$_2\times$S$^1$ vacuum, there is a warped (A)dS \cite{Arvanitakis:2014yja},
a solution that describes gravitational waves \cite{Alishahiha:2014dma}, a two-parameter
deformation of BTZ black hole \cite{Giribet:2014wla} and a non-BTZ AdS black hole 
\cite{Arvanitakis:2015yya}. It has been shown that all solutions of TMG that have Petrov-types
O, N and D are also solutions of MMG after a proper redefinition of parameters \cite{Altas:2015dfa}.
In \cite{Altas:2015dfa} it is also found that MMG inherits from TMG a specific type of Kundt solution.

In this article we construct and algebraically classify all Kundt solutions, i.e. spacetimes with 
an expansion-free null geodesic congruence, of MMG theory. Kundt spacetimes in three dimensions
are quite special: They are known to be primary examples of spacetimes for which all scalar 
invariants constructed from the relevant curvature tensors and their covariant derivatives are
constants. All such, so-called ``constant scalar invariant" (CSI), spacetimes in three dimensions
have been classified in \cite{Coley:2005sq}: Apart from Kundt solutions, there are only locally 
homogeneous geometries with this property. In our analysis to determine these special solutions,
we will adapt the strategy followed by \cite{Chow:2009vt} to find the analogous solutions of TMG.
We find that as TMG, the CSI Kundt solutions of MMG turn out to be deformations of round and 
warped (A)dS. However, not all Kundt solutions are CSI and indeed we also find an exact 
non-CSI Kundt solution for a specific fine-tuning of two of the four MMG parameters. Then, we 
algebraically classify the solutions we find with respect to their traceless Ricci tensor, i.e. present 
their algebraic Segre-types, making use of the results of \cite{Chow:2009km}. It turns out that their 
Segre-types match with those of their relevant counterparts in TMG. In passing, it is worthy of noting 
that the Kundt solutions of NMG were studied in \cite{Chakhad:2009em}, where a non-CSI Kundt 
solution to TMG was also found at the chiral point \cite{Li:2008dq}.

The paper is organised as follows: In section \ref{model} we briefly review MMG theory. Section
\ref{Kundt} is devoted to the detailed derivation of the Kundt solutions obtained and forms the
bulk of our paper. The algebraic Segre classification of the Kundt solutions found are given in 
section \ref{algcla}. We conclude and discuss possible future problems in section \ref{disc}. In
two separate appendices, we give certain technical details that we left out in the main text. 

\section{The MMG model}\label{model}
MMG theory itself is closely related to TMG \cite{Deser:1982vy, Deser:1981wh}, whose source-free 
field equation reads
\begin{equation}
R_{\mu\nu} - \frac{1}{2} R \, g_{\mu\nu} + \Lambda_{0} \, g_{\mu\nu} + \frac{1}{\mu} C_{\mu\nu} = 0 \,,
\label{TMG}
\end{equation}
where the symmetric, traceless, parity-odd and covariantly conserved Cotton tensor $C_{\mu\nu}$ 
is defined in terms of the Schouten tensor $S_{\sigma\nu}$ as
\begin{equation}
C^{\mu}\,_{\nu} \equiv \epsilon^{\mu\rho\sigma} \, \nabla_{\rho} \, S_{\sigma\nu}  \,, \qquad 
S_{\sigma\nu} \equiv R_{\sigma\nu} - \frac{1}{4} R \, g_{\sigma\nu} \,,
\label{cotten}
\end{equation}
and the Levi-Civita pseudo tensor is defined as
\( \epsilon_{\mu\rho\sigma} = \sqrt{-g} \, \varepsilon_{\mu\rho\sigma} \)
in terms of the weight +1 tensor density \( \varepsilon_{\mu\rho\sigma} \), where we use 
the convention \( \varepsilon_{012} = +1 \). Here $\Lambda_{0}$ is the cosmological constant,
and $\mu$ is a mass parameter (with dimensions 1/Length).

On the other hand, the field equation of source-free MMG theory reads
\begin{equation}
\bar{\sigma} \Big( R_{\mu\nu} - \frac{1}{2} R \, g_{\mu\nu} \Big) + \bar{\Lambda}_{0} \, g_{\mu\nu} 
+ \frac{1}{\mu} C_{\mu\nu} + \frac{\gamma}{\mu^2} J_{\mu\nu} = 0 \,,
\label{MMG}
\end{equation}
where the symmetric, curvature-squared tensor $J_{\mu\nu}$ is
\begin{eqnarray}
J^{\mu\nu} & \equiv & - \frac{1}{2} \, \epsilon^{\mu\rho\sigma} \, \epsilon^{\nu\tau\eta} \, 
S_{\rho\tau} \, S_{\sigma\eta} \,, \nonumber \\
& = & S^{\mu\rho} S^{\nu}\,_{\rho} - S^{\mu\nu} S 
- \frac{1}{2} g^{\mu\nu} \Big( S^{\rho\sigma} S_{\rho\sigma} - S^2 \Big) \,, \label{jten} \\
& = & R^{\mu\rho} R^{\nu}\,_{\rho} - \frac{3}{4} R^{\mu\nu} R 
- \frac{1}{2} g^{\mu\nu} \Big( R^{\rho\sigma} R_{\rho\sigma} - \frac{5}{8} R^2 \Big) \,, \nonumber
\end{eqnarray}
with \( S \equiv g^{\mu\nu} S_{\mu\nu} \). Even though the $J$-tensor is not covariantly conserved
on its own, i.e. it does not satisfy a Bianchi identity so that the MMG field equation (\ref{MMG})
can not be derived from an action that only involves the metric and its curvature tensors, it is
nevertheless conserved \emph{on-shell} as a consequence of (\ref{MMG}) itself (see 
\cite{Bergshoeff:2014pca, Arvanitakis:2014yja} for details). Moreover, it is possible to construct 
conserved charges for solutions of the theory \cite{Tekin:2014jna}. In (\ref{MMG}), the parameters 
$\bar{\sigma}$ and $\gamma$ are non-zero dimensionless constants, whereas 
$\bar{\Lambda}_{0}$ is the cosmological constant with dimensions 1/Length$^2$. Note that 
the trace of (\ref{MMG}) implies that
\begin{equation}
\bar{\sigma} R - 6 \bar{\Lambda}_{0} 
+ \frac{\gamma}{\mu^2} \Big( R^{\mu\nu} R_{\mu\nu} - \frac{3}{8} R^2 \Big) = 0 \, .
\label{trMMG}
\end{equation}

The parameters showing up in MMG field equation can be expressed in terms of those 
of TMG \cite{Bergshoeff:2014pca} as\footnote{To be able to make smooth contact with 
the Kundt solutions of TMG \cite{Chow:2009vt} in the limit \( \gamma \to 0 \), we made
the choice \( \sigma = 1 \) in these formulas.}:
\begin{eqnarray}
\bar{\sigma} & = & 1 + \alpha + \frac{\alpha^2 \Lambda_0}{2 \mu^2 (1 + \alpha)^2} \,, \nonumber \\
\gamma & = & - \frac{\alpha}{(1+\alpha)^2} \,, \label{parameters} \\
\bar{\Lambda}_0 & = & \Lambda_0 \left(1 + \alpha - \frac{\alpha^3 \Lambda_0}{4 \mu^2 (1 + \alpha)^2} \right) \,, \nonumber 
\end{eqnarray}
where $\alpha$ is a dimensionless parameter such that one gets TMG in the \( \alpha \to 0 \) limit.
For bulk and boundary unitarity it is necessary to have \cite{Arvanitakis:2014xna}:
\begin{equation}
-1 < \alpha < 0 \;, \qquad \Lambda_0  <  \frac{4 \mu^2 (1 + \alpha)^3}{\alpha^3} \,. \label{unitarity}
\end{equation}
Note that these conditions imply \( \Lambda_0 <0 \) and \( \Lambda_0 \alpha > 0 \). (See also the
equations (5.13) and (5.15), and the discussion that leads to them, of \cite{Bergshoeff:2014pca}.)

There are two special points in the parameter space of the MMG theory. The first one is called the 
`chiral point' \cite{Li:2008dq} for which the central charges vanish and is given by 
\cite{Bergshoeff:2014pca}, \cite{Tekin:2014jna}:
\begin{equation}
 {\bar \sigma} + \frac{\gamma}{2\mu^2 \ell^2_{ch.}} \pm \frac{1}{\mu \ell_{ch.}} =0 \,,  \qquad \qquad
 \frac{1}{\mu^2 \ell^2_{ch.}} \equiv {\bar \sigma}^2 - \frac{\gamma \bar{\Lambda}_0}{\mu^2} \,.
 \label{chiral}
\end{equation}
The second one is called the `merger point' \cite{Arvanitakis:2014yja} where: 
\begin{equation}
\bar{\Lambda}_{0} = \frac{\mu^2 \bar{\sigma}^2}{\gamma} \, . 
\label{merger}
\end{equation}
For this choice the quadratic equation for the effective cosmological constant of maximally 
symmetric vacua has a repeated root. (See \cite{Arvanitakis:2015yya} for further discussion.) 
Note that the use of (\ref{parameters}) in the condition (\ref{merger}) implies that
\begin{equation}
\Lambda_0 \alpha = - \mu^2 (1 + \alpha)^2 < 0 \,,
\end{equation}
which violates the unitarity conditions (\ref{unitarity}), since it demands \( \Lambda_0 \alpha > 0 \).

\section{Kundt solutions}\label{Kundt}
MMG field equation (\ref{MMG}) is highly nonlinear and involves fourth order derivatives, and 
we now want to look for Kundt solutions of the theory. To facilitate the comparison of possible 
Kundt solutions with the known Kundt solutions of TMG found in \cite{Chow:2009vt}, we 
hence use the same strategy, the same notations and conventions used there.

Thus, taking the orientation convention \( \epsilon_{vu\rho} = 1 \), we start with
a general Kundt spacetime given by the metric
\begin{equation}
ds^2 = d\rho^2 + 2 \, du \, dv + f(v, u, \rho) \, du^2 + 2 \, W(v, u, \rho) \, du \, d\rho \,.
\label{met}
\end{equation}
Note that the null vector required in the definition of a Kundt spacetime is 
\( k^{\mu} = \delta^{\mu}_{\;\;v} \), which makes \( k_{\mu} = \delta_{\mu\,u} \), and
fulfils the following conditions:
\begin{eqnarray}
(a) & & \; \mbox{expansion free:} \quad \nabla_{\mu} k^{\mu} = 0 \,, \nonumber \\
(b) & & \; \mbox{shear free:} \quad (\nabla^{\mu} k^{\nu}) \nabla_{(\mu} k_{\nu)} 
- (\nabla_{\mu} k^{\mu})^2 = 0 \,, \label{defK} \\
(c) & & \; \mbox{twist free:} \quad (\partial^{\mu} k^{\nu}) \partial_{[\mu} k_{\nu]} = 0 \, .\nonumber 
\end{eqnarray}

The $vv$-component of the field equation (\ref{MMG}) gives
\begin{equation}
\gamma \Big( \frac{\partial^2 W}{\partial v^2} \Big)^2 + 2 \mu \frac{\partial^3 W}{\partial v^3} = 0 \,. 
\label{vvcomp}
\end{equation}
The most general solution of (\ref{vvcomp}) is
\begin{equation} 
W(v, u, \rho) = \frac{2 \mu}{\gamma^2} \Big( -v \gamma + \big( v \gamma - 2 \, \mu \, W_2(u, \rho) \big)
 \ln{\big( v \gamma - 2 \, \mu \, W_2(u, \rho) \big)} \Big) + v \, W_1(u, \rho) + W_0(u, \rho) \,. 
\label{Wsol}
\end{equation}
As shown in detail in appendix \ref{appa}, unless one takes \( v \gamma - 2 \mu W_2(u, \rho) = 0 \),
one ends up in a dead end. Hence, as argued for the TMG case in \cite{Chow:2009vt} 
(and see also \cite{Podolsky:2008ec} for the analogous result on Kundt solutions in higher 
dimensions), we continue with a metric function $W$ which is linear in $v$ rather than 
using the most general solution. Thus we take
\begin{equation} 
W(v, u, \rho) = v \, W_1(u, \rho) + W_0(u, \rho) \,. 
\label{Wsollin}
\end{equation}
Now the $v\rho$-component of the field equation becomes\footnote{See (\ref{vrhocomp}) 
of Appendix \ref{appa} for its form in the generic case.}
\begin{equation} 
\frac{\partial^3 f}{\partial v^3} = 0 \,,
\label{nvrhocomp}
\end{equation}
which is easily integrated as
\begin{equation} 
f(v, u, \rho) = v^2 \, f_2(u, \rho) + v \, f_1(u, \rho) + f_0(u, \rho) \,. 
\label{fsollin}
\end{equation}

The form of the metric (\ref{met}) (with metric functions $W$ and $f$ as in (\ref{Wsollin}) and (\ref{fsollin}), respectively) is left invariant under the following coordinate 
transformations\footnote{Here we have directly reproduced equation (2.9) of \cite{Chow:2009vt}. 
Refer to equations (2.9) to (2.11) of \cite{Chow:2009vt} for details.}
\begin{equation}
v = \frac{\tilde{v}}{\dot{u}(\tilde{u})} + F(\tilde{u}, \tilde{\rho}) \,, \quad
u = u(\tilde{u}) \,, \quad \dot{u}(\tilde{u}) \equiv \frac{d u}{d \tilde{u}} \,, \quad
\rho = \tilde{\rho} + G(\tilde{u}) \,.
\label{coor}
\end{equation}
These will be needed in the discussion that follows.

Using (\ref{Wsollin}) and (\ref{fsollin}) in (\ref{trMMG}) and a bit of massaging, one now obtains
\begin{equation}
\frac{\gamma}{4 \mu^2} \Big( f_2 - \frac{1}{4} W_1^2 \Big) 
\Big( 2 \frac{\partial W_1}{\partial \rho} - f_2 - \frac{3}{4} W_1^2 \Big)
- \bar{\sigma} \Big( \frac{\partial W_1}{\partial \rho} + f_2 - \frac{3}{4} W_1^2 \Big)
+ 3 \bar{\Lambda}_{0} = 0 \,,
\label{trlin}
\end{equation}
a quadratic equation for $f_2$ unlike the case for TMG. One can easily solve it to write
\begin{eqnarray}
f_2(u, \rho) & = & 
- \frac{2 \mu^2 \bar{\sigma}}{\gamma} + \frac{\partial W_1}{\partial \rho} - \frac{1}{4} W_1^2 \nonumber \\
& & \pm \sqrt{\frac{12 \mu^2}{\gamma} 
\Big( \frac{\mu^2 \bar{\sigma}^2}{3 \gamma} + \bar{\Lambda}_{0} \Big) 
+ \Big( \frac{\partial W_1}{\partial \rho} - \frac{1}{2} W_1^2 \Big) 
\Big( \frac{\partial W_1}{\partial \rho} - \frac{1}{2} W_1^2 - \frac{8 \mu^2 \bar{\sigma}}{\gamma} \Big)} \,,
\label{f2sol}
\end{eqnarray}
which makes the $\gamma \to 0$ limit, for comparison with the TMG case, intractable at first sight. 
It turns out that the trace equation can be written as a linear combination of the $vu$ and 
$\rho\rho$-components of the field equation\footnote{The trace equation (\ref{trlin}) equals 
the sum of the $\rho\rho$- and twice the $vu$-component of the field equation.}. Concentrating 
only on the $\rho\rho$-component, one can write it as
\begin{equation}
- \frac{1}{\mu} \frac{\partial f_2}{\partial \rho} + \frac{1}{\mu} f_2 W_1 
+ \frac{\gamma}{4 \mu^2} \Big( f_2 - \frac{1}{4} W_1^2 \Big)^2
- \bar{\sigma} \Big( f_2 - \frac{1}{4} W_1^2 \Big) + \bar{\Lambda}_{0} = 0 \, .
\label{rrcomp}
\end{equation}
Substituting (\ref{f2sol}) into (\ref{rrcomp}), one obtains
\begin{eqnarray}
- \frac{1}{\mu} \frac{\partial}{\partial \rho} \left[ \chi \pm \sqrt{\Big( \chi - \frac{4 \mu^2 \bar{\sigma}}{\gamma} \Big)^2 + \frac{12 \mu^2}{\gamma} 
\Big( \bar{\Lambda}_{0} - \frac{\mu^2 \bar{\sigma}^2}{\gamma} \Big) } \, \right] 
+ \bar{\Lambda}_{0} - \frac{\mu^2 \bar{\sigma}^2}{\gamma} & & \nonumber \\
+ \frac{\gamma}{4 \mu^2} \left[ - \frac{4 \mu^2 \bar{\sigma}}{\gamma} 
+ \chi \pm \sqrt{\Big( \chi - \frac{4 \mu^2 \bar{\sigma}}{\gamma} \Big)^2 + \frac{12 \mu^2}{\gamma} 
\Big( \bar{\Lambda}_{0} - \frac{\mu^2 \bar{\sigma}^2}{\gamma} \Big) } \, \right]^2 & & \label{W1eqn} \\
+ \frac{W_1}{\mu}  \left[ - \frac{2 \mu^2 \bar{\sigma}}{\gamma} 
+ \frac{1}{2} \chi \pm 
\sqrt{\Big( \chi - \frac{4 \mu^2 \bar{\sigma}}{\gamma} \Big)^2 + \frac{12 \mu^2}{\gamma} 
\Big( \bar{\Lambda}_{0} - \frac{\mu^2 \bar{\sigma}^2}{\gamma} \Big) } \, \right] & = & 0 \,, \nonumber 
\end{eqnarray}
where
\begin{equation}
\chi(u, \rho) \equiv \frac{\partial W_1}{\partial \rho} - \frac{1}{2} W_1^2 \,.
\label{chidef}
\end{equation}
In principle, one can determine the function $W_1(u, \rho)$ from (\ref{W1eqn}). If one has
a solution which is only $\rho$-dependent, say $W_1(\rho)$, then for an arbitrary smooth 
enough function $c(u)$, $W_1(\rho + c(u))$ will also be a solution because of (\ref{coor}). 
So, out of the two ``integration constants" present in the solution of (\ref{W1eqn}), one can 
be discarded since the transformation \( \rho \to \rho - c(u) \) can be employed to that effect
\footnote{This immediately follows from the first equation in (2.11) of \cite{Chow:2009vt}.} 
for a generic $\rho$-dependent $W_1$. Hence, in what follows we will take 
\( W_1 = W_1(\rho) \) only.

Once $W_1(\rho)$ is determined from (\ref{W1eqn}), the coordinate transformation 
\( v \to v + F(u, \rho) \) can be employed to set $W_0 = 0$. Then the 
$u\rho$-component of the field equation turns out to be of the form
\begin{equation}
A \Big( W_1, \frac{dW_1}{d\rho} \Big) \, v 
+ B \Big( W_1, \frac{dW_1}{d\rho}, \frac{\partial f_1}{\partial \rho}, \frac{\partial^2 f_1}{\partial \rho^2} \Big) = 0 \,,
\label{urhocomp}
\end{equation}
where the coefficient functions $A$ and $B$, with their indicated arguments, are rather
long and complicated. Unlike the TMG case, where the coefficient function $A$ identically 
vanishes, both $A$ and $B$ must equal zero for (\ref{urhocomp}) to hold. For a generic 
$W_1(\rho)$ however, it turns out that the differential constraints (\ref{W1eqn}) and $A=0$ 
together are too strong and yield an overdetermined system with only the trivial solution. 

Hence we trace back the steps we have taken from (\ref{trlin}) to (\ref{W1eqn}) and ask whether 
there are any ``simple" choices one can make to find some ``special" classes of solutions. 
Thus we are led to the following cases to consider:
\begin{eqnarray}
\mbox{A)} & \quad & \chi(\rho) \equiv \frac{d W_1}{d \rho} - \frac{1}{2} W_1^2 = w_1 = \mbox{const.} 
\,, \label{choiA} \\
\mbox{B)} & \quad & W_1 = \omega = \mbox{const.} \,, \label{choiB} \\
\mbox{C)} & \quad & \bar{\Lambda}_{0} = \frac{\mu^2 \bar{\sigma}^2}{\gamma} \;\;
(\mbox{merger point})\, \,. \label{choiD} 
\end{eqnarray}
We will show that the first two choices lead to CSI solutions, whereas there is a non-CSI solution for
the last one. Let us also note that even though the case \( f_2 - W_1^2/4 = \mbox{const.} \) seems 
to be special at first sight, as we show in appendix \ref{Xcon}, it turns out that this only leads to 
special instances of solutions obtained in other cases, which we now study in detail below.

\subsection{\(\chi  = \mbox{const.}\)}\label{chicon}
As argued after (\ref{chidef}), taking \( W_1 = W_1(\rho) \), one easily integrates (\ref{choiA})
to find\footnote{Here we again use the transformation \( \rho \to \rho - \mbox{const.} \) to 
drop an integration constant.}
\begin{equation}
W_1(\rho) = \sqrt{2 w_1} \, \tan{\Big( \rho \, \sqrt{ \frac{w_1}{2} } \Big)} \,.
\label{W1sol}
\end{equation}
The solution of $f_2$, that follows from (\ref{trlin}) and replaces (\ref{f2sol}), is
\begin{equation}
f_2(u, \rho) = - \frac{2 \mu^2 \bar{\sigma}}{\gamma} 
+ \frac{w_1}{2} \left[ 1+ \sec^2{\Big( \rho \, \sqrt{ \frac{w_1}{2} } \Big)} \right]  
\pm \sqrt{ \Big( w_1 - \frac{4 \mu^2 \bar{\sigma}}{\gamma} \Big)^2 
+ \frac{12 \mu^2}{\gamma}\Big( \bar{\Lambda}_{0} - \frac{\mu^2 \bar{\sigma}^2}{\gamma} \Big)} \,.
\label{f2solA}
\end{equation}
Note that the $u$-dependency of $f_2$ has effectively dropped altogether.
Substituting (\ref{W1sol}) and (\ref{f2solA}) into the $\rho\rho$-component of the field equation
(\ref{rrcomp}), one finds a long expression which is a linear function of $W_1(\rho)$. Setting
the coefficient of $W_1(\rho)$ to zero, which can be shown to be equivalent to the condition  
\begin{equation}
\gamma w_1^2 - 8 \mu^2 \bar{\sigma} w_1 + 16 \mu^2 \bar{\Lambda}_{0} = 0 \,, \label{w1eqn}
\end{equation}
one can fix the constant $w_1$ uniquely, depending on the sign in the solution of 
$f_2$ (\ref{f2solA}), as\footnote{Note that $\omega_1$ is twice the value of the allowed 
cosmological constant for maximally symmetric vacua of MMG, found in equation (2.3) 
of \cite{Arvanitakis:2014yja}. (\ref{w1solA}) simplifies considerably at the chiral point (\ref{chiral}) 
and the merger point (\ref{merger}).}
\begin{equation}
w_1 = \frac{4 \mu^2 \bar{\sigma}}{\gamma} \Big( 1 
\mp \sqrt{1- \frac{\gamma \bar{\Lambda}_{0}}{\mu^2 \bar{\sigma}^2}} \, \Big) \,.
\label{w1solA}
\end{equation}
Substituting (\ref{w1solA}) back into (\ref{f2solA}), one finds a much simpler expression for $f_2$:
\begin{equation}
f_2(u, \rho) = \frac{1}{4} (W_1(\rho))^2 + \frac{w_1}{2} 
= \frac{w_1}{2} \sec^2{\Big( \rho \, \sqrt{ \frac{w_1}{2} } \Big)} \,.  \label{f2simp}
\end{equation}
Setting \( f_1(u,\rho) = 0 \) and \( f_0(u,\rho) = 0 \), one finds the ``background metric", which
is itself a solution to MMG, as
\begin{equation}
ds^2 = d\rho^2 + 2 \, du \, dv + \frac{w_1}{2} \sec^2{\Big( \rho \, \sqrt{ \frac{w_1}{2} } \Big)} \, v^2 \, du^2
+ 2 \, v \, \sqrt{2 w_1} \, \tan{\Big( \rho \, \sqrt{ \frac{w_1}{2} } \Big)} \, du \, d\rho \,,
\label{metback}
\end{equation}
with curvature scalar \( R = 3 w_1 \) and \( R^{\mu\nu} R_{\mu\nu} = 3 w_1^2 \).

Turning \( f_1(u,\rho) \) and \( f_0(u,\rho) \) on, the $u\rho$-component of the field equation 
reads\footnote{For clarity of argument, $w_1$ found in (\ref{w1solA}) has not been used in the 
remainder of this subsection.}
\begin{equation}
\frac{\partial^2 f_1}{\partial \rho^2} + 
\left[ \mu \bar{\sigma} - \frac{w_1 \gamma}{4 \mu} - \sqrt{\frac{w_1}{2}} \, 
\tan{\Big( \rho \, \sqrt{ \frac{w_1}{2} } \Big)} \right] \frac{\partial f_1}{\partial \rho} = 0 \,.
\label{f1eqn}
\end{equation}
This can be thought of as a first order linear partial differential equation for $\partial f_1/\partial \rho$,
that can easily be solved as
\[ \frac{\partial f_1}{\partial \rho} = \, f_{11}(u) \, \E^{a \rho} \, \sec{\Big(\rho\,\sqrt{ \frac{w_1}{2} } \Big)} 
\,, \quad a \equiv \frac{w_1 \gamma}{4 \mu} - \mu \bar{\sigma} \,. \]
Thus the most general solution of (\ref{f1eqn}) reads
\begin{equation}
f_1(u, \rho) = F(\rho) \, f_{11}(u) + f_{12}(u) \,, \quad
F(\rho) \equiv \int^{\rho} d\tilde{\rho} \, \E^{a \tilde{\rho}} \, \sec{\Big(\tilde{\rho} \,\sqrt{ \frac{w_1}{2} } \Big)} \,.
\label{f1sol} 
\end{equation}
A coordinate transformation of the form \( v \to v/(du/d\tilde{u}) \,, u \to u(\tilde{u}) \), can 
be employed to set \( f_{12}(u) = 0 \)\footnote{See (2.11) of \cite{Chow:2009vt} for details.}. 
Finally the $uu$-component of the field equation reads
\begin{eqnarray}
\frac{\partial^3 f_0}{\partial \rho^3} +  \Big( \frac{3}{2} W_1(\rho) 
+ \mu \bar{\sigma} - \frac{w_1 \gamma}{4 \mu}\Big) 
\left[ \frac{\partial^2 f_0}{\partial \rho^2} + W_1(\rho) \frac{\partial f_0}{\partial \rho} 
+ \frac{1}{2} (W_1(\rho))^2 f_0(u, \rho) \right] & \nonumber \\
+ 2 w_1 \frac{\partial f_0}{\partial \rho} + \Big( \frac{3}{2} w_1 W_1(\rho) 
+ \mu \big( 4 \bar{\Lambda}_{0} - w_1 \bar{\sigma} \big) \Big) f_0(u, \rho) & & \label{f0eqn} \\
= \frac{d F}{d \rho} \left[ \frac{(f_{11}(u))^2}{2}  
\Big( \frac{\gamma}{\mu} \frac{d F}{d \rho} - F(\rho) \Big) - \frac{d f_{11}}{d u} \right] \,, \nonumber
\end{eqnarray}
which is a linear partial differential equation for the remaining metric function $f_0(u, \rho)$. Given
$f_{11}(u)$, one can in principle solve it to find $f_0(u, \rho)$. Hence the generic Kundt
solution for this case is
\begin{eqnarray}
ds^2 & = & d\rho^2 + 2 \, du \, dv 
+ \left[ \frac{w_1}{2} \sec^2{\Big( \rho \, \sqrt{ \frac{w_1}{2} } \Big)} \, v^2 + f_1(u, \rho) \, v 
+ f_0(u, \rho) \right] \, du^2 \nonumber \\
& & + 2 \, v \, \sqrt{2 w_1} \, \tan{\Big( \rho \, \sqrt{ \frac{w_1}{2} } \Big)} \, du \, d\rho \,,
\label{metgen}
\end{eqnarray}
with a real $w_1$ determined as in (\ref{w1solA}), $f_1(u,\rho)$ as given by (\ref{f1sol}) (with
$f_{11}(u)$ arbitrary and \( f_{12}(u) = 0 \)) and $f_0(u, \rho)$ satisfying (\ref{f0eqn}).

Inspired by the discussion given in subsection 4.3.2 of \cite{Chow:2009vt}, if one defines 
a new coordinate 
\[ \hat{v} = u - \frac{4 \cos^2{\big( \rho \, \sqrt{w_1/2} \big)}}{w_1 v} \,, \]
then the solution (\ref{metback}) takes the form
\begin{equation}
ds^2 = d\rho^2 + \frac{8 \cos^2{\big( \rho \, \sqrt{w_1/2} \big)}}{w_1 (u - \hat{v})^2} \, du \, d\hat{v} \,,
\label{metbacknew}
\end{equation}
which is the metric for the round (A)dS. Thus (\ref{metgen}) can be thought of as the deformation
of the round (A)dS metric, cast in the generalised Kerr-Schild form with an (A)dS background.

\subsection{\(W_1 = \mbox{const.}\)}\label{w1con}
When one sets \( W_1 = \omega = \mbox{const.} \), (\ref{trlin}) can be arranged to read
\begin{equation}
f_2^2 + f_2 \Big( \frac{\omega^2}{2} + \frac{4 \mu^2 \bar{\sigma}}{\gamma} \Big)
- 3 \Big( \frac{\omega^4}{16} + \frac{\mu^2}{\gamma} 
\big( \bar{\sigma} \omega^2 + 4 \bar{\Lambda}_{0}\big) \Big) = 0 \,,
\label{f2eqn1}
\end{equation}
which clearly implies that $f_2$ must also be a constant. Using this in the $\rho\rho$-component 
of the field equation (\ref{rrcomp}), one finds that this can be written as
\begin{equation}
f_2^2 - f_2 \Big( \frac{\omega^2}{2} + \frac{4 \mu}{\gamma} \big( \mu \bar{\sigma} - \omega \big) \Big)
+ \Big( \frac{\omega^4}{16} + \frac{\mu^2}{\gamma} 
\big( \bar{\sigma} \omega^2 + 4 \bar{\Lambda}_{0}\big) \Big) = 0 \,.
\label{f2eqn2}
\end{equation}
Since (\ref{f2eqn1}) and (\ref{f2eqn2}) must be satisfied simultaneously, one can simply solve
for $\bar{\Lambda}_{0}$ using either one of these and use this in the other to solve for
$f_2$. Doing so, one finds that there are two possibilities:
\begin{equation}
f_2 = 0 \qquad \mbox{or} \qquad
f_2 = \frac{\omega^2}{4} - \frac{3 \mu \omega}{\gamma} + \frac{2 \mu^2 \bar{\sigma}}{\gamma} \,.
\label{f2spe}
\end{equation}
If one sets $f_2=0$ in either (\ref{f2eqn1}) or (\ref{f2eqn2}), one finds that these are satisfied only if
\begin{equation}
\omega^2 = - \frac{8 \mu^2 \bar{\sigma}}{\gamma} \Big( 1 
\mp \sqrt{1- \frac{\gamma \bar{\Lambda}_{0}}{\mu^2 \bar{\sigma}^2}} \, \Big) \,,
\label{omesol}
\end{equation}
which simplifies both at the chiral point (\ref{chiral}) and the merger point (\ref{merger}). 
At this stage if one were to further set \( f_1(u,\rho) = 0 \) and \( f_0(u,\rho) = 0 \), one would
simply obtain the spacetime of constant curvature
\begin{equation}
ds^2 = d\rho^2 + 2 \, du \, dv + 2 \, \omega \, v \, du \, d\rho \,,
\label{met1}
\end{equation}
with curvature scalar \( R = - 3 \omega^2/2 \) and \( R^{\mu\nu} R_{\mu\nu} = 3 \omega^4/4 \). 
This, of course, is a solution to MMG with any real $\omega$ satisfying (\ref{omesol}).

However, one can dress up this ``background metric" further in the following way:
The $u\rho$-component of the field equation yields\footnote{For clarity, neither of the 
roots in (\ref{omesol}) has been used explicitly in the remainder of this subsection.}
\begin{equation}
\frac{\partial^2 f_1}{\partial \rho^2} + 
\left[ \mu \bar{\sigma} + \frac{\gamma \omega^2}{8 \mu} - \frac{\omega}{2} \right] 
\frac{\partial f_1}{\partial \rho} = 0 \,.
\label{f1eqnom}
\end{equation}
Using a coordinate transformation (similar to the one employed after (\ref{f1sol}) in subsection 
\ref{chicon}) to set an arbitrary function of $u$ to zero, we take the solution of (\ref{f1eqnom}) as
\begin{equation}
f_1(u, \rho) = F(\rho) \, f_{11}(u) \,, \quad \mbox{where} \;\; F(\rho) = \E^{- \big( \mu \bar{\sigma} + 
\frac{\gamma \omega^2}{8 \mu} - \frac{\omega}{2} \big) \rho} \,, \label{f1solom}
\end{equation}
with $f_{11}(u)$ an arbitrary function\footnote{A careful scrutiny of equations (2.9) to (2.11) of 
\cite{Chow:2009vt} shows that the function $f_{11}(u)$ cannot be ``gauged away" by the 
coordinate transformations (\ref{coor}).}. Then the $uu$-component of the field equation is
\begin{equation}
\frac{\partial^3 f_0}{\partial \rho^3} + \Big( \frac{3 \omega}{2} 
+ \mu \bar{\sigma} + \frac{\gamma \omega^2}{8 \mu} \Big) \left[ \frac{\partial^2 f_0}{\partial \rho^2}
+ \omega \frac{\partial f_0}{\partial \rho} \right] - \omega^2 \frac{\partial f_0}{\partial \rho}
= \frac{d F}{d \rho} \left[ \frac{(f_{11}(u))^2}{2}  
\Big( \frac{\gamma}{\mu} \frac{d F}{d \rho} - F(\rho) \Big) - \frac{d f_{11}}{d u} \right] \,, \label{f0eqnom}
\end{equation}
a linear partial differential equation for the metric function $f_0(u, \rho)$. As in subsection \ref{chicon},
one can in principle solve it to find $f_0(u, \rho)$ given $f_{11}(u)$. Putting it all together, the 
Kundt solution found is
\begin{equation}
ds^2 = d\rho^2 + 2 \, du \, dv + 2 \, \omega \, v \, du \, d\rho 
+ \big( v \, f_1(u, \rho) + f_0(u, \rho) \big) \, du^2 \,, \label{met1dress}
\end{equation}
where $\omega$ is any one of the real roots of (\ref{omesol}), $f_1(u,\rho)$ is given by 
(\ref{f1solom}) with $f_{11}(u)$ arbitrary, and $f_0(u, \rho)$ satisfies (\ref{f0eqnom}).

Note that the coordinate transformation \( \hat{v} = v \E^{\omega \rho} \) turns (\ref{met1}) into
\begin{equation}
ds^2 = d\rho^2 + 2 \E^{-\omega \rho} \, du \, d\hat{v} \,,
\label{met1new}
\end{equation}
which is simply the round AdS \cite{Chow:2009vt}. Thus (\ref{met1dress}) can be considered 
as a deformation of the round AdS, cast again in a generalised Kerr-Schild form with an AdS 
background.

Now one can also use the other choice of $f_2$ in (\ref{f2spe}), i.e. that
\begin{equation}
f_2 = \frac{\omega^2}{4} - \frac{3 \mu \omega}{\gamma} + \frac{2 \mu^2 \bar{\sigma}}{\gamma} \,.
\label{f2spe2}
\end{equation}
Before proceeding any further, note that this can also be written as\footnote{This alternative 
form will be of use for the discussion in appendix \ref{Xcon}.}
\begin{equation}
f_2 = \frac{\omega^2}{4} + \xi \,, \quad \mbox{where} \quad 
\xi \equiv - \frac{3 \mu \omega}{\gamma} + \frac{2 \mu^2 \bar{\sigma}}{\gamma} \quad \mbox{or}
\quad \omega = \frac{2 \mu \bar{\sigma}}{3} - \frac{\gamma \xi}{3 \mu} \,.
\label{f2spealt}
\end{equation}
In this case both (\ref{trlin}) and (\ref{rrcomp}) imply
\begin{equation}
\frac{\omega^3}{4 \mu} - \frac{3 \omega^2}{4 \gamma} + \frac{2 \mu \bar{\sigma} \omega}{\gamma}
+ \big( \bar{\Lambda}_{0} - \frac{\mu^2 \bar{\sigma}^2}{\gamma} \big) = 0 \,,
\label{rrtreq}
\end{equation}
or for easier access to the \( \gamma \to 0 \) limit
\begin{equation}
\frac{\gamma^3}{108 \mu^4} \xi^3 + \frac{\gamma}{6 \mu^2} \Big( \frac{1}{2} 
- \frac{\gamma \bar{\sigma}}{3} \Big) \xi^2 + \frac{\bar{\sigma}}{3} 
\Big( 1 + \frac{\gamma \bar{\sigma}}{3} \Big) \xi - 
\Big( \bar{\Lambda}_{0} + \frac{2 \mu^2 \bar{\sigma}^3}{27} \Big) = 0 \,,
\label{xieqn}
\end{equation}
a cubic equation in $\omega$ (or $\xi$), the explicit roots of which are better left not displayed. 
One can now ``play the game" of setting \( f_1(u,\rho) = 0 \) and \( f_0(u,\rho) = 0 \), and 
obtain the background metric
\begin{equation}
ds^2 = d\rho^2 + 2 \, du \, dv + 2 \, \omega \, v \, du \, d\rho
+ \Big( \frac{\omega^2}{4} - \frac{3 \mu \omega}{\gamma} + \frac{2 \mu^2 \bar{\sigma}}{\gamma} \Big) 
\, v^2 \, du^2 \,,
\label{metspe}
\end{equation}
which has a curvature scalar
\[ R = - \omega^2 + 2 \xi \quad \mbox{and} \quad 
R^{\mu\nu} R_{\mu\nu} = \frac{1}{8} \big( 2 \omega^4 + (\omega^2 - 4 \xi)^2 \big) \,, \]
with $\omega$ any one of the real root(s) of (\ref{rrtreq}).

Turning $f_1$ and $f_0$ on is straightforward: The $u\rho$-component of the field equation is
\begin{equation}
\frac{\partial^2 f_1}{\partial \rho^2} + \omega \frac{\partial f_1}{\partial \rho} = 0 \,,
\label{f1eqspe}
\end{equation}
which is readily integrated, with an arbitrary function \( f_{12}(u) = 0 \) as before, as
\begin{equation}
f_1(u, \rho) = \E^{- \omega \rho} \, f_{11}(u) \,. \label{f1solspe}
\end{equation}
However, the coordinate transformation
\[ v \to v - \frac{1}{2 f_2} \E^{- \omega \rho} \, f_{11}(u) \]
lets one take \( f_{11}(u) = 0 \), so that one can set \( f_1(u,\rho) = 0 \) for good. Finally, the 
$uu$-component of the field equation becomes
\begin{equation}
\frac{\partial^3 f_0}{\partial \rho^3} 
+ \Big( 2 \mu \bar{\sigma} + \frac{\gamma \omega^2}{4 \mu} \Big) \frac{\partial^2 f_0}{\partial \rho^2}
- \Big( \frac{\omega^2}{2} - \frac{3 \mu \omega}{\gamma} + \frac{2 \mu^2 \bar{\sigma}}{\gamma}
+ \gamma \bar{\Lambda}_{0} - \mu^2 \bar{\sigma}^2 \Big) \frac{\partial f_0}{\partial \rho} = 0 \,, 
\label{f0eqnspe}
\end{equation}
whose most general solution is easy to obtain:
\begin{eqnarray}
f_0(u, \rho) & = & f_{01}(u) + f_{02}(u) \E^{- \big( \Omega + 3 \omega + \gamma f_2/\mu \big) \rho/2} 
+ f_{03}(u) \E^{\big( \Omega - 3 \omega - \gamma f_2/\mu \big) \rho/2} \,, \label{f0solspe} \\
\mbox{with} \quad \Omega & \equiv & \sqrt{10 \omega^2 + \frac{6 \gamma f_2 \omega}{\mu} + 
\frac{\gamma^2 f_2^2}{\mu^2} + 4 (f_2 + \gamma \bar{\Lambda}_{0} - \mu^2 \bar{\sigma}^2)} \,, 
\nonumber
\end{eqnarray}
for arbitrary functions $f_{01}(u)$, $f_{02}(u)$ and $f_{03}(u)$ with $f_2$ given in (\ref{f2spe2}). 
Thus, with \( f_0(u, \rho) \) determined as in (\ref{f0solspe}), the generic Kundt solution for this case is 
\begin{equation}
ds^2 = d\rho^2 + 2 \, du \, dv + 2 \, \omega \, v \, du \, d\rho
+ \left[ \Big( \frac{\omega^2}{4} - \frac{3 \mu \omega}{\gamma} + \frac{2 \mu^2 \bar{\sigma}}{\gamma} 
\Big) \, v^2 + f_0(u, \rho) \right] \, du^2 \,.
\label{metspefull}
\end{equation}
A specific solution of this type with \( f_0 = 1 \) has already been found in \cite{Altas:2015dfa}.

Note first that the metric (\ref{metspe}) can be written in the simple form
\begin{equation}
ds^2 = 2 \, du \, dv + \big( d\rho + \omega \, v \, du \big)^2 - q \, v^2 \, du^2 \,,
\label{metspe1}
\end{equation}
where 
\[ q \equiv \omega^2 - f_2 = \frac{3 \omega^2}{4} + \frac{3 \mu \omega}{\gamma} - \frac{2 \mu^2 
\bar{\sigma}}{\gamma} \,. \]
The coordinate transformations \( \hat{u} = q u/2 \) and \( \hat{v} = 1/v + q u/2 \) takes (\ref{metspe1}) into the form
\begin{equation}
ds^2 = - \frac{4 \, d\hat{u} \, d\hat{v}}{q (\hat{u} - \hat{v})^2} + 
\Big( d\rho - \frac{2 \, \omega \, d\hat{u}}{q (\hat{u} - \hat{v})} \Big)^2 \,.
\label{metspe2}
\end{equation}
Finally the coordinates \( t = (\hat{u} + \hat{v})/2 \), \( x = (\hat{v} - \hat{u})/2 \) and 
\( z = q \rho/\omega -\ln{(\hat{v} - \hat{u})} \) can be employed to cast (\ref{metspe2}) into
\begin{equation}
ds^2 = \frac{1}{q} \Big[ \frac{-dt^2 +dx^2}{x^2} + 
\frac{\omega^2}{q} \Big( dz + \frac{dt}{x} \Big)^2 \Big] \,,
\label{metspe3}
\end{equation}
which renders this solution as the spacelike-warped (A)dS \cite{Chow:2009vt, Arvanitakis:2014yja}. 
Thus (\ref{metspefull}) can be thought of as the deformation of spacelike-warped (A)dS, written 
in a generalised Kerr-Schild form with a spacelike-warped (A)dS background.

\subsection{Merger Point: \(\bar{\Lambda}_{0} = \mu^2 \bar{\sigma}^2/\gamma\)}\label{spe}
For this special fine-tuning of the parameters the equation (\ref{trlin})
factorizes:
\begin{equation}
\frac{\gamma}{4 \mu^2} \Big( f_2 - \frac{1}{4} W_1^2 - \frac{2 \mu^2 \bar{\sigma}}{\gamma} \Big) 
\Big( f_2 - 2 \frac{d W_1}{d \rho} + \frac{3}{4} W_1^2 
+ \frac{6 \mu^2 \bar{\sigma}}{\gamma} \Big) = 0 \,, \label{trspe}
\end{equation}
which lets $f_2$ to be determined in terms of $W_1$. 

The substitution of the first choice
\begin{equation}
f_2 = \frac{1}{4} W_1^2 + \frac{2 \mu^2 \bar{\sigma}}{\gamma} \label{f21st}
\end{equation}
into the $\rho\rho$-component of the field equation (\ref{rrcomp}) gives
\begin{equation}
W_1 \Big( \frac{d W_1}{d \rho} - \frac{1}{2} W_1^2 - \frac{4 \mu^2 \bar{\sigma}}{\gamma} \Big) = 0 \,.
\label{W11st}
\end{equation}
If one takes \( W_1 = 0 \), so that \( f_2 = 2 \mu^2 \bar{\sigma}/\gamma \), then one ends up
finding the solution described in the second half of subsection \ref{w1con} with \( \omega = 0 \).
If, however, (using our earlier notation of (\ref{choiA})) one takes 
\( \chi = 4 \mu^2 \bar{\sigma}/\gamma = \mbox{const.} \), one is led to the 
general solution described in subsection \ref{chicon} with \( w_1 = 4 \mu^2 \bar{\sigma}/\gamma \).

The substitution of the second choice, namely
\begin{equation}
f_2 = 2 \frac{d W_1}{d \rho} - \frac{3}{4} W_1^2 - \frac{6 \mu^2 \bar{\sigma}}{\gamma} \,, \label{f22nd}
\end{equation}
into the $\rho\rho$-component of the field equation (\ref{rrcomp}) is a bit subtler. One finds
\begin{equation}
\frac{d}{d\rho} \Big( \frac{d W_1}{d \rho} - \frac{1}{2} W_1^2 - \frac{4 \mu^2 \bar{\sigma}}{\gamma} \Big) 
- \frac{\gamma}{2 \mu} \Big( \frac{d W_1}{d \rho} - \frac{1}{2} W_1^2 - \frac{4 \mu^2 \bar{\sigma}}
{\gamma} \Big)^2 - \frac{3}{4} W_1 \Big( \frac{d W_1}{d \rho} - \frac{1}{2} W_1^2 - \frac{4 \mu^2 
\bar{\sigma}}{\gamma} \Big) = 0 \,, \label{eq33}
\end{equation}
a \emph{nonlinear} second-order ordinary differential equation from which $W_1(\rho)$ can 
in principle be solved. It is easy to see that the particular choices, the cases \( W_1 = 0 \) and 
\( \chi = 4 \mu^2 \bar{\sigma}/\gamma \) that we have already covered, solve it. 
Now we present a `particular' solution of (\ref{eq33}) which is a non-CSI spacetime as we will see.
Provided that out of the three remaining `free' 
parameters, $\bar{\sigma}, \mu$ and $\gamma$, of the MMG theory, the pair $\bar{\sigma}$
and $\gamma$ are further fine-tuned as \( 2 \bar{\sigma} \gamma + 1 = 0 \), such that
\( \bar{\Lambda}_{0} = \mu^2 /(4 \gamma^3) \) now, one finds that (\ref{eq33}) is identically
satisfied if
\begin{equation}
\frac{d W_1}{d \rho} - \frac{1}{2} W_1^2 + \frac{\mu}{2 \gamma} W_1 + \frac{\mu^2}{\gamma^2} = 0 \,.
\label{neweq33}
\end{equation}
Employing the transformation \( \rho \to \rho - \mbox{const.} \) to get rid of an integration constant
as previously argued, the solution of this equation reads
\begin{equation}
W_1(\rho) = \frac{\mu}{\gamma} \, \Big( \frac{2 + \E^{3 \mu \rho/(2 \gamma)}}{1-\E^{3 \mu \rho/(2 \gamma)}} \Big)
\,, \label{nonCSIW1}
\end{equation}
which determines $f_2$ through (\ref{f22nd}) as
\begin{equation}
f_2 = \Big( \frac{1}{2} W_1 - \frac{\mu}{\gamma} \Big)^2 = \frac{9 \mu^2}{4 \gamma^2} \, 
\frac{\E^{3 \mu \rho/\gamma}}{\big( 1- \E^{3 \mu \rho/(2 \gamma)} \big)^2} \,. \label{nonCSIf2}
\end{equation}
Thus one arrives at the background metric
\begin{equation}
ds^2 = d\rho^2 + 2 \, du \, dv + \frac{9 \mu^2}{4 \gamma^2} \, 
\frac{\E^{3 \mu \rho/\gamma}}{\big(1- \E^{3 \mu \rho/(2 \gamma)} \big)^2} \, v^2 \, du^2 
+ 2 \, \frac{\mu}{\gamma} \, 
\Big( \frac{2 + \E^{3 \mu \rho/(2 \gamma)}}{1-\E^{3 \mu \rho/(2 \gamma)}} \Big) \, \, v \, du \, d\rho \,,
\label{nonCSImetback}
\end{equation}
with $f_1$ and $f_0$ switched off. This metric has curvature scalar
\[ R = - \frac{3 \mu^2}{\gamma^2} \, \Big( \frac{2 + \E^{3 \mu \rho/(2 \gamma)}}{1-\E^{3 \mu \rho/(2 \gamma)}} \Big)
\quad \mbox{and} \quad 
R^{\mu\nu} R_{\mu\nu} = \frac{3 \mu^4}{8 \gamma^4} \, 
\frac{\big( 32 + 32 \E^{3 \mu \rho/(2 \gamma)} + 17 \E^{3 \mu \rho/\gamma}\big)}
{\big( 1-\E^{3 \mu \rho/(2 \gamma)} \big)^2} \,. \]
Note that there is a curvature singularity at $\rho=0$ and as $\rho \to \infty$ the metric 
(\ref{nonCSImetback}) approaches to the CSI metric given in (\ref{metspe}) with 
\( \omega=-\mu/\gamma\) (and hence \( \xi = 2 \mu^2/\gamma^2 \)), which is a root 
of (\ref{rrtreq}), at this special setting. Note that the asymptotic CSI metric has 
positive curvature scalar \( R = 3 \mu^2/\gamma^2 > 0 \) then.

Turning $f_1$ and $f_0$ on is straightforward. One finds that $f_1$ must satisfy
\begin{equation}
\frac{\partial^2 f_1}{\partial \rho^2} - \frac{\mu}{\gamma}
\frac{\partial f_1}{\partial \rho} = 0 \,, \label{nonCSIf1eq}
\end{equation}
which is easily integrated, with an arbitrary function \( f_{12}(u) = 0 \) as before, as
\begin{equation}
f_1(u, \rho) = \E^{\mu \rho/\gamma} \, f_{11}(u) \,. \label{nonCSIf1sol}
\end{equation}
On the other hand, the coordinate transformation
\[ v \to v - \frac{1}{2 f_2} \E^{\mu \rho/\gamma} \, f_{11}(u) \]
can be employed to set \( f_{11}(u) = 0 \), so that one has \( f_1(u,\rho) = 0 \) for good.
The equation that $f_0$ must satisfy is
\begin{eqnarray}
\frac{\partial^3 f_0}{\partial \rho^3} +  \frac{3 \mu}{4 \gamma} \, \Big( \frac{4 + \E^{3 \mu \rho/(2 
\gamma)}}{1-\E^{3 \mu \rho/(2 \gamma)}} \Big) \frac{\partial^2 f_0}{\partial \rho^2} +
\frac{\mu^2}{4 \gamma^2} \, 
\frac{\big( 8 + 47 \E^{3 \mu \rho/(2 \gamma)} - 10 \E^{3 \mu \rho/\gamma}\big)}
{\big( 1-\E^{3 \mu \rho/(2 \gamma)} \big)^2} \frac{\partial f_0}{\partial \rho} & \nonumber \\ +
\frac{9 \mu^3}{8 \gamma^3} \, \frac{\E^{3 \mu \rho/(2 \gamma)} \big( 14 + \E^{3 \mu \rho/(2 \gamma)} 
\big)}{\big( 1-\E^{3 \mu \rho/(2 \gamma)} \big)^3} f_0
& = 0 \,. \label{nonCSIf0eqn} 
\end{eqnarray}
Thus the general Kundt solution for this case is
\begin{equation}
ds^2 = d\rho^2 + 2 \, du \, dv + \Big[ \frac{9 \mu^2}{4 \gamma^2} \, 
\frac{\E^{3 \mu \rho/\gamma}}{\big(1- \E^{3 \mu \rho/(2 \gamma)} \big)^2} \, v^2 + f_0(u, \rho) \Big] \, du^2 + 2 \, \frac{\mu}{\gamma} \, 
\Big( \frac{2 + \E^{3 \mu \rho/(2 \gamma)}}{1-\E^{3 \mu \rho/(2 \gamma)}} \Big) \, \, v \, du \, d\rho \,,
\label{nonCSImetgen}
\end{equation}
where it is understood that $f_0$ satisfies the linear partial differential equation (\ref{nonCSIf0eqn}).

\section{Algebraic Classification}\label{algcla}
In three dimensions the algebraic classification of curvature can be achieved by using either  
the traceless Ricci tensor 
\[ \hat{R}^{\mu}\,_{\nu} \equiv R^{\mu}\,_{\nu} - \frac{1}{3} R \delta^{\mu}\,_{\nu} \]
(Segre classification) or the Cotton tensor (\ref{cotten}) (Petrov classification). Even though these 
two classifications coincide for TMG, this is not so for MMG. Since the Cotton tensor involves 
the derivative of the Ricci tensor, and thus can be considered as ``less fundamental", we 
concentrate on the traceless Ricci tensor for the algebraic classification of the Kundt solutions 
of MMG. In this case, one must thus find the eigenvalues of \( \hat{R}^{\mu}\,_{\nu} \) and 
their multiplicities, if any. Since the Jordan normal form encodes all of this information, we 
calculate this for each of the solutions obtained thus far. However it will be convenient to first 
recall the basics of the Segre notation we use: The symbols 1, 2 and 3 indicate the sizes of 
Jordan blocks. Parentheses are used for grouping Jordan blocks that belong to the same 
eigenvalue. A comma is used for splitting the Jordan blocks corresponding to spacelike 
eigenvectors from those corresponding to the timelike ones, where the former are always written 
before the comma. (Refer to table 1 of \cite{Chow:2009km} for details.)

A generic Kundt spacetime (\ref{met}), where $W$ and $f$ are given as in (\ref{Wsollin}) 
and (\ref{fsollin}), respectively, is of Segre-type $[12]$, where the eigenvalues of 
\( \hat{R}^{\mu}\,_{\nu} \) are $\beta$ and $-2 \beta$ with algebraic multiplicities 2 and 1, 
respectively, and 
\begin{equation}
\beta \equiv \frac{1}{6} \Big( 2 f_2 - \frac{\partial W_1}{\partial \rho} \Big) \,.
\label{eigval}
\end{equation}

For the solutions found in subsection \ref{chicon}, it turns out that $\beta = 0$. The solution
(\ref{metback}) is of Segre-type $[(12)]$, whereas (\ref{metgen}) is of Segre-type $[3]$.
Since \(W_1 = \mbox{const.}\) and $f_2$ must be as in (\ref{f2spe}) for the solutions 
found in subsection \ref{w1con}, there are two possibilities for each choice of $f_2$:
For the trivial case $f_2 = 0$, it follows that $\beta = 0$ as well. Then the solution
(\ref{met1}) is of Segre-type $[(11,1)]$, whereas (\ref{met1dress}) is of Segre-type $[3]$. 
For the non-trivial $f_2$ given as in (\ref{f2spe2}), it is obvious that $\beta = f_2/3 \neq 0$ 
and it turns out that the solution (\ref{metspe}) is of Segre-type $[1(1,1)]$, whereas 
the solution (\ref{metspefull}) is of Segre-type $[12]$. Note that solutions presented in 
appendix \ref{Xcon}, namely (\ref{met2}), and its dressed-up version (\ref{met3}) are both 
closely related to their counterparts (\ref{metspe}) and (\ref{metspefull}) of subsection 
\ref{w1con}, respectively, and are of the same Segre-type as their cousins. Finally, for the 
non-CSI solution presented in subsection \ref{spe}, namely (\ref{nonCSImetback}) and its 
sister with $f_0$ turned on (\ref{nonCSImetgen}), the specific value of the eigenvalue is
\[ \beta = - \frac{3 \mu^2}{4 \gamma^2} \, \Big( \frac{\E^{3 \mu \rho/(2 \gamma)}}{1-\E^{3 \mu \rho/(2 \gamma)}} \Big) \,, \]
and both of these solutions are again of Segre-type $[12]$.

Comparing the Segre-types of these solutions to the Segre-types of the Kundt solutions of
TMG, one sees that the presence of the parameter $\gamma$ does not alter the general
picture. The Kundt solutions of MMG also fall into two broad classes of Segre-type $[12]$
and Segre-type $[3]$, with special cases of Segre-types $[(11,1)]$, $[1(1,1)]$ and $[(12)]$
also occurring.

\section{Discussion}\label{disc}
In this paper we investigated the Kundt solutions of MMG theory. All of the explicit solutions
presented in subsections \ref{chicon} and \ref{w1con} turn out to be CSI spacetimes, which are 
deformations of round and warped (A)dS. We also showed the existence of Kundt solutions at the 
merger point (\ref{merger}) in subsection \ref{spe} and found an explicit non-CSI solution. Even 
though the fine-tuned parameters of the latter violate the unitarity conditions (\ref{unitarity}), there 
aren't any restrictions on the parameters of the former and they remain indifferent to the unitarity 
requirements. The chiral point (\ref{chiral}) does not seem to be special for Kundt solutions except for 
simplifying relevant expressions. Since the algebraic classification is a useful tool for identifying 
seemingly different spacetimes, we also gave the Segre classification of these solutions and found that 
their Segre-types coincide with their cousins in TMG.

There are various possible directions to extend our work. First of all, one may try to relax 
the expansion-free condition on the null vector $k^{\mu}$ in (\ref{defK}) and look for 
Robinson-Trautman solutions. One may also try to find all locally homogeneous solutions 
as was done for TMG \cite{Moutsopoulos:2012bi, Siampos:2013foa}. This is an interesting 
problem, since then all CSI solutions of MMG can be completed. Another open question is 
whether the Goldberg-Sachs theorem is also valid for MMG. This was recently proven for 
TMG in \cite{Nurowski:2015hwa}. Studying such solutions in models that are related
to MMG would also be interesting \cite{Setare:2014zea, Tekin:2015rha}. We hope to 
investigate these and related issues in the near future.

\begin{acknowledgments}
NSD and {\"O}S are partially supported by the Scientific and Technological Research 
Council of Turkey (T{\"U}B\.{I}TAK) grant 113F034.
\end{acknowledgments}

\appendix
\section{\label{appa} The most general solution of (\ref{vvcomp}):}
As argued in the beginning of section \ref{Kundt}, here we show that the Kundt spacetimes 
exist only if one takes \( Y(v, u, \rho) \equiv v \gamma - 2 \mu W_2(u, \rho) = 0 \).

Recall that the most general solution of (\ref{vvcomp}) is given by (\ref{Wsol}).
Using (\ref{vvcomp}), the $v\rho$-component of the field equation (\ref{MMG}) reads
\begin{equation} 
\left[ 8 \mu^2 \bar{\sigma} + \gamma \Big( \frac{\partial W}{\partial v} \Big)^2 
- 2 \gamma \frac{\partial^2 f}{\partial v^2} \right] \frac{\partial^2 W}{\partial v^2} - 4 \mu \frac{\partial^3 f}{\partial v^3} = 0 \,,
\label{vrhocomp}
\end{equation}
which after the substitution of (\ref{Wsol}) can be cast into 
\begin{equation}
8 \gamma \mu^2 \bar{\sigma} +
\Big( \gamma W_1(u, \rho) + 2 \mu \ln{\big( v \gamma - 2 \, \mu \, W_2(u, \rho) \big)} \Big)^2 
- 2 \gamma  \frac{\partial }{\partial v} \Big( \big( v \gamma - 2 \, \mu \, W_2(u, \rho) \big) 
\frac{\partial^2 f}{\partial v^2} \Big) = 0 \,.
\label{feqn}
\end{equation}
Assuming that \( Y(v, u, \rho) \equiv v \gamma - 2 \, \mu \, W_2(u, \rho) \neq 0 \), this partial differential 
equation for the metric function $f$ can be solved to determine its $v$ dependency as
\begin{eqnarray}
f(v, u, \rho) & = & v^2 \left\{ \frac{\mu^2}{2 \gamma^2} 
\left[ 17 + 4 \gamma \bar{\sigma} - \Big(10 - 2 \ln{Y} \Big)
\ln{Y} \right] - \frac{\mu}{2 \gamma} W_1(u, \rho)
\Big( 5 - 2 \ln{Y} \Big)
+ \frac{1}{4} W_1^2(u, \rho) \right\} \nonumber \\
& & + v \left\{ f_1(u, \rho) 
- \frac{f_2(u, \rho)}{\gamma} \Big(1 - \ln{Y} \Big) 
- \frac{2 \mu^3}{\gamma^3} W_2(u, \rho) 
\left[17 + 8 \gamma \bar{\sigma} - \Big(10 + 4 \gamma \bar{\sigma} 
- 2 \ln{Y} \Big) \ln{Y} \right] \right. \nonumber \\
& & \left. \qquad + \frac{2 \mu^2}{\gamma^2} W_1(u, \rho) W_2(u, \rho)
\Big( 7 - 4 \ln{Y} \Big) - \frac{\mu}{\gamma} W_1^2(u, \rho) W_2(u, \rho)
\Big( 2 - \ln{Y} \Big) \right\} \label{fsol} \\
& & + f_0(u, \rho) - \frac{2 \mu}{\gamma^2} f_2(u, \rho) W_2(u, \rho) 
\ln{Y} - \frac{2 \mu^4}{\gamma^4} W_2^2(u, \rho) \left[ 5 + 2 \ln{Y} 
\Big( 5 + 4 \gamma \bar{\sigma} - \ln{Y} \Big) \right] \nonumber \\
& & - \frac{2 \mu^3}{\gamma^3} W_1(u, \rho) W_2^2(u, \rho)
\Big( 1 - 6 \ln{Y} \Big) - \frac{2 \mu^2}{\gamma^2} W_1^2(u, \rho) W_2^2(u, \rho)
\ln{Y} \,. \nonumber
\end{eqnarray}

Using (\ref{Wsol}) and (\ref{fsol}) in (\ref{trMMG}), one obtains a very long expression,
better not displayed here at all, whose general form is
\[ \sum_{m=0}^{2} \sum_{n=0}^{3} a_{mn}(u, \rho) \, v^{m} \,
(\ln{\big( v \gamma - 2 \, \mu \, W_2(u, \rho) \big)})^{n} = 0 \,. \]
Obviously each of the coefficients $a_{mn}(u, \rho)$ must vanish for this equality to
hold, but the vanishing of the ``highest" one, \( a_{23}(u, \rho) = 96 \gamma^2 \mu^4 = 0 \), 
implies that either $\gamma = 0$ or $\mu = 0$, both of which takes us out of MMG theory.

\section{\label{Xcon} The case \( f_2 - W_1^2/4 = \mbox{const.} \) in (\ref{trlin})}
In this appendix, we study what happens when one chooses 
\( f_2 - W_1^2/4 = \mbox{const.} \) in equation (\ref{trlin}) and afterwards. Dropping the 
$u$-dependency of $f_2$ and setting \( f_2(\rho) - W_1(\rho)^2/4 = \xi = \mbox{const.} \), 
(\ref{trlin}) simplifies to
\begin{equation} 
\xi (\gamma \xi + 4 \mu^2 \bar{\sigma}) - 12 \mu^2 \bar{\Lambda}_{0} 
- 2 \chi(\rho) (\gamma \xi - 2 \mu^2 \bar{\sigma}) = 0 \,, \label{Xconxi}
\end{equation}
where we have used the definition of $\chi(\rho)$ as given in (\ref{choiA}), and taken
\( W_1 = W_1(\rho) \) as argued after (\ref{chidef}). The generic solution of 
(\ref{Xconxi}) falls precisely into the \( \chi = \mbox{const.} \) case already discussed in 
detail in subsection \ref{chicon}\footnote{The field equations then impose the 
identification of the constant $w_1$ of subsection \ref{chicon} as $w_1 = 2 \xi$.}. 

Note also that if one were to set \( \xi=0 \) to start with, and solve for the relevant version 
of (\ref{Xconxi}) keeping \( W_1 = W_1(\rho) \) generic, then one immediately finds that 
(\ref{rrcomp}) can only be satisfied if \( \bar{\Lambda}_{0} = 0 \). This at once leads to the 
``trivial" solution: Both $W_1$ and $f_2$ must vanish and one simply obtains the flat
Minkowski metric
\begin{equation}
ds^2 = d\rho^2 + 2 \, du \, dv \,, \label{mink}
\end{equation}
when one keeps both $f_1$ and $f_0$ switched off. When one chooses to turn $f_1$ and 
$f_0$ on, the analysis flows exactly as in the second paragraph of subsection \ref{w1con} 
with the substitution $\omega=0$.

Still keeping $\xi=0$, if one instead takes \( W_1 = \tilde{\omega} = \mbox{const.} \), one 
finds from (\ref{trlin}) that\footnote{Of course, the constraint \( \bar{\Lambda}_{0}/\bar{\sigma} < 0 \) 
(or \( \bar{\Lambda}_{0} \bar{\sigma} < 0 \)) is implicitly assumed in what follows.} 
\( \tilde{\omega}^2 = - 6 \bar{\Lambda}_{0}/\bar{\sigma} \), which makes 
\( f_2 = - 3 \bar{\Lambda}_{0}/(2 \bar{\sigma}) \). Substituting these in (\ref{rrcomp}), one 
ends up with either \( \bar{\Lambda}_{0} = 0 \) or 
\( \bar{\Lambda}_{0} = - 2 \mu^2 \bar{\sigma}^3/27 \). The former case goes ahead as
mentioned in the previous paragraph. The latter gives \( W_1 = 2 \mu \bar{\sigma}/3 \) and 
\( f_2 = \mu^2 \bar{\sigma}^2/9 \)\footnote{Note that these choices correspond to taking 
$\xi=0$ in (\ref{f2spealt}) and (\ref{xieqn}).}. If one chooses to set \( f_1(u,\rho) = 0 \) and 
\( f_0(u,\rho) = 0 \) here, then one simply obtains the spacetime of constant negative curvature
\begin{equation}
ds^2 = d\rho^2 + 2 \, du \, dv + \frac{\mu^2 \bar{\sigma}^2}{9} \, v^2 \, du^2
+ \frac{4 \mu \bar{\sigma}}{3} \, v \, du \, d\rho \,,
\label{met2}
\end{equation}
with \( R = -4 \mu^2 \bar{\sigma}^2/9 \) and \( R^{\mu\nu} R_{\mu\nu} = 2 \mu^4 \bar{\sigma}^4/27 \), 
as a solution to MMG. The story with the turning on of the other functions $f_1$ and $f_0$ follows 
similar steps as before, but we skip the details for clarity and simply present the final result:
The general Kundt solution for this case is\footnote{This reduces to the metric given in equation 
(3.4) of \cite{Chow:2009vt} in the limit \( \gamma \to 0 \), \( \bar{\sigma} \to 1 \) and 
\( \Lambda \to - 2 \mu^2/27 \).}
\begin{equation}
ds^2 = d\rho^2 + 2 \, du \, dv + \Big( \frac{\mu^2 \bar{\sigma}^2}{9} \, v^2 + f_0(u, \rho) \Big) \, du^2
+ \frac{4 \mu \bar{\sigma}}{3} \, v \, du \, d\rho \,,
\label{met3}
\end{equation}
where a coordinate transformation as in (\ref{coor}) has been employed on $v$ to set 
\( f_1(u,\rho) = 0 \) for good and $f_0(u, \rho)$ reads
\begin{eqnarray}
f_0(u, \rho) & = & f_{01}(u) 
+ f_{02}(u) \E^{- \mu \bar{\sigma} \big( 18 + \gamma \bar{\sigma} + \Xi \big) \rho/18} 
+ f_{03}(u) \E^{- \mu \bar{\sigma} \big( 18 + \gamma \bar{\sigma} - \Xi \big) \rho/18} \,, \label{f0sollam} \\
\mbox{with} \quad \Xi & \equiv & \sqrt{\gamma^2 \bar{\sigma}^2 + 12 \gamma \bar{\sigma} +72} \,, 
\nonumber
\end{eqnarray}
for arbitrary functions $f_{01}(u)$, $f_{02}(u)$ and $f_{03}(u)$.

To cover all bases, if one were to examine the assumption \( W_1(\rho) = \bar{\omega} 
= \mbox{const.} \) with $\xi \neq 0$, then one soon finds that this reduces to the case
covered in the first half of subsection \ref{w1con}, with the identification 
\( \bar{\omega} \to \omega \) and $f_2 = 0$, etc.

Finally, if one were to set the coefficient of $\chi(\rho)$ in (\ref{Xconxi}) to zero, i.e. set
\( \xi = 2 \mu^2 \bar{\sigma}/\gamma \), then the remainder of (\ref{Xconxi}) demands 
\( \bar{\Lambda}_{0} = \mu^2 \bar{\sigma}^2/\gamma \) which is examined in detail
in subsection \ref{spe}.

\end{document}